\documentstyle[aps,preprint]{revtex}
\topmargin = - 0.9 cm
\begin{document}
\begin{titlepage}
\begin{flushright}
IFUP-TH-12/95
\end{flushright}
\vskip 1truecm
\begin{center}
\Large\bf
Faddeev--Popov determinant in \\
2--dimensional Regge gravity\footnote{This work is  supported in part
  by M.U.R.S.T.}
\end{center}

\vskip 1truecm
\begin{center}
{Pietro Menotti and Pier Paolo Peirano} \\
{\small\it Dipartimento di Fisica dell' Universit\`a, Pisa 56100,
Italy and}\\
{\small\it INFN, Sezione di Pisa}\\
\end{center}
\vskip .8truecm
\begin{center}
March 1995
\end{center}
\end{titlepage}

\begin{abstract}
\noindent
By regularizing the singularities appearing in the two dimensional
Regge calculus by means of a segment of a sphere or pseudo-sphere and
then taking the regulator to zero, we obtain a simple
formula for the gauge volume which appears in the functional integral.
Such a formula is an analytic function of the opening of the conic
singularity  in the interval from $\pi$ to $4\pi$ and in the continuum
limit it goes over to the correct result.

\end{abstract}

Much interest has been recently devoted to the discrete approach to
two dimensional gravity both in the dynamical triangulation
and in the Regge approach \cite{amb,hamb}.  It
appears that dealing with such a problem in the Regge framework
needs a correct treatment of the measure \cite{jan}.  The
attitude we shall adopt here \cite{lee} is that to consider the
Regge skeleton as
defining a geometry described by a metric $g_{\mu\nu}$, which is flat
except at the vertices where the curvature becomes singular.

The functional integration has to be performed on the metric,
with a distance defined by the De Witt super-metric. In order to do
that, a gauge fixing has to be introduced and one has to keep into
account the correct gauge volume given by the Faddeev--Popov (FP)
determinant. Such an approach is the one which has proven successful
in the continuum formulation \cite{poly,alva,moor} and has
been advocated e.g.\ by Jevicki and Ninomiya \cite{jev} in the Regge
framework.  In practice the functional integration is given by the
integral over the link lengths $l_i$ multiplied by a determinant which
takes into account how the gauge fixed metric depends on the $l_i$,
times the FP determinant. The computation of the first determinant is
rather straightforward as it does not involve divergent
quantities. Instead we shall concentrate here on the computation of
the FP part. It will be shown that following a procedure developed by
Aurell and Salomonson \cite{aur} for the computation of
the determinant of the scalar Laplacian, it is possible to give an
exact expression for such a determinant.  In the continuum limit it
will reproduce the well known result.  The FP determinant in question
is given by ${\det}'(L^{\dag}L)$ where $L$ is the differential operator
which takes from a vector field $\xi_\mu$ which generates a
diffeomorphism to the traceless part of the change in the metric and
$L^{\dag}$ is its adjoint. In general if we denote by $\hat
g_{\mu\nu}$ the background metric and with $\sigma$ the conformal
factor with $g=e^{2\sigma}\hat g$, we have $L=e^{2\sigma} \hat L
e^{-2\sigma}$ and $L^{\dag}=e^{-D\sigma} \hat L^{\dag}
e^{(D-2)\sigma}$ where $D$ is the dimension of space-time.  In two
dimensions we shall use the complex coordinate $\omega$ and write
$\xi=\xi_1+i\xi_2$ for the vector field and $h=h_{11} +i h_{12}$ for
the traceless symmetric tensor field.  With the background metric
$\hat g_{\mu\nu} = \delta_{\mu\nu}$, $\hat L$ and $\hat L^{\dag}$ take
the simple form $\displaystyle{\hat L={\partial\over\partial \bar
\omega}}$ and $\displaystyle{\hat L^{\dag}=-{\partial\over \partial
\omega}}$ and thus
\begin{equation}
L^{\dag}L =- e^{-2\sigma}{\partial\over \partial \omega}e^{2 \sigma}
{\partial\over \partial \bar{\omega}}e^{-2\sigma}
\end{equation}
and
\begin{equation}
LL^{\dag} = - e^{2\sigma}{\partial\over \partial \bar{\omega}}e^{-4 \sigma}
{\partial\over \partial \omega}.
\end{equation}
${\det}' (L^{\dag}L)$ is a divergent quantity and as usual it will be defined
through the procedure of $Z$-function regularization. We have
\begin{equation}
Z(s)={1\over \Gamma(s)}\int_0^\infty dt~ t^{s-1} {\rm Tr}'(e^{-t L^{\dag}L})
\end{equation}
and
\begin{equation}
-\log({\det}'(L^{\dag}L))= Z'(s)|_{s=0} = \gamma_E Z(0) + {\rm
Finite_{\epsilon \rightarrow 0}}
\int^\infty_\epsilon {dt\over t} {\rm Tr}'(e^{-tL^{\dag}L})
\end{equation}
where det$'$ denotes the determinant after removing the zero modes.

Following the standard procedure developed in the continuum approach,
$Z'(0)$ will be computed by first performing a variation $\delta\sigma$
in the conformal factor and later integrating back the result.
For the case at hand we have
\begin{eqnarray}
 \delta Z'(0) = \gamma_E\delta c^{K}_0 +  {\rm Finite_{\epsilon
\rightarrow 0}} \int d^2 x [ 4
\sigma(x) (K(x,x,\epsilon) -\sum_{\nu} |\Psi_{\nu}
(x) |^{2}) +   & \nonumber \\
 - 2 \delta   \sigma(x) (H(x,x,\epsilon) -
\sum_{\nu} |\Phi_{\nu} (x) |^{2})]  & \label{var}
\end{eqnarray}
where $K$ is the heat kernel of the operator $L^{\dag}L$, $H$ is the
heat kernel of the operator $LL^{\dag}$;  $\Psi_{\nu}(x)$  and
$\Phi_{\nu}(x)$ denote the normalized zero modes of $L^{\dag}L$ and
$LL^{\dag}$ respectively and $c^{K}_0 = Z(0) + {\rm dim
Ker}~(L^{\dag}L) $ is the constant term in the
asymptotic expansion of the trace of $K(x,x',t)$.

The main point is the computation of $K(x,x',t)$ and $H(x,x',t)$ which
respect the correct boundary conditions imposed by the nature of the
vector field $\xi$ and of the tensor field $h$.  The cone can be
described in the $z=x+iy$ plane, by a wedge of angular opening
$2\pi\alpha$ ($\alpha =1$ is the plane).  The conformal variation
considered is the same adopted
in \cite{aur} for the scalar Laplacian, which takes from a
cone with some opening angle $2\pi \alpha$
and extension $\lambda$ to a cone with varied opening angle $\alpha
+\delta\alpha$ and extension $\lambda +\delta\lambda$.  Such a
conformal transformation is described by the variation \cite{aur} in
the $z$-plane
\begin{equation}
\delta \sigma(z,\bar z) =(\delta \lambda - \lambda
{\delta \alpha\over \alpha})
+ {\delta \alpha\over \alpha} \log(\alpha|z|).
\label{delta}
\end{equation}
The vector field $\xi_\mu$ that describes the 2-dimensional
diffeomorphisms on a cone  can be represented by  the complex scalar
$\xi$ provided one imposes that the phase change $2\pi\delta$ on the boundary,
equals the opening angle $2\pi\alpha$ of the cone modulo
$2\pi$. A similar condition holds for the complex field $h$
representing the tensor, provided we replace $2\pi\delta$ by
$4\pi\delta$ modulo $2\pi$.

The spectral representation of the heat kernel on the cone is given
by
\begin{eqnarray}
K_{\alpha,\delta}(r,r';t) = \frac{1}{2\pi\alpha} \left\{
   \sum_{n=0}^{\infty} e^{i (\phi -\phi') \frac{n+\delta}{\alpha}}
   \int_{0}^{\infty} J_{\frac{n+\delta}{\alpha}}(r\mu)
   J_{\frac{n+\delta}{\alpha}} (r'\mu) e^{-t\mu^{2}} \, \mu d
   \mu +    \right. & \nonumber \\
 + \left.  \sum_{n=1}^{\infty} e^{-i (\phi -\phi') \frac{n-\delta}{\alpha}}
          \int_{0}^{\infty} J_{\frac{n-\delta}{\alpha}}(r\mu)
          J_{\frac{n-\delta}{\alpha}} (r'\mu) e^{-t\mu^{2}} \, \mu d
          \mu   \right\}  &  \label{green3}
\end{eqnarray}
and is valid for Re $\nu > -1$ ($\nu$ is the index of the Bessel
function). Following a well known procedure \cite{cars,dow}
eq.(\ref{green3}) can be rewritten as a contour integral
\begin{equation}
K_{\alpha,\delta}(r,r';t) = \frac{1}{16 i \pi^2\alpha t} \int_{A} \!
d\zeta \;
e^{-\frac{1}{4t} (r^2 + r'^2 - 2 r r' \cos \zeta)} \; \;
\frac{e^{\frac{i}{2\alpha}(\zeta + \phi - \phi')(2\delta -
1)}}{\sin \frac{\zeta + \phi - \phi' }{2\alpha}}  \label{green1}
\end{equation}
where the contour $A$ is composed of two branches, one starting at
$\pi + i \infty$ and ending at $-\pi + i \infty$ and the other
starting at $-\pi - i \infty$ and ending at $\pi -i \infty$.
One can obtain the constant term $c^K_0$ in the asymptotic
expansion of the heat kernel (\ref{green1}), by first extracting
the contribution on the plane and, after taking the trace inside the
integral, by a proper deformation of the integration contour. We remark
that such a procedure works for $|2\delta -1| < 2 \alpha + 1$.
The result is \cite{dow,chee}
\begin{equation}
\displaystyle c^K_0= \frac{\delta(\delta - 1)}{2\alpha} +
\frac{1-\alpha^{2}}{12\alpha }.
\label{dow}
\end{equation}
Dowker's choice for $\delta$ is $0<\delta \leq 1$ and gives rise to a heat
kernel which is regular at the origin.  This is simply checked by
showing that for $r=0$, $K$ vanishes because in this case one can
close the integration contours at infinity. This is not the case for
$\delta <0$ or $\delta >1$, as can be seen from eq.(\ref{green3}) due
to the appearance of Bessel's functions of negative index, even if the
integral for $r$ and $r'$ different from zero still converges.

A straightforward application of (\ref{dow}) with the standard choice
$0 < \delta \leq 1$ gives for $\alpha=1 + \varepsilon$ with
$\varepsilon < 0$,
$\displaystyle c^K_0 = \frac{\varepsilon}{3} + O(\varepsilon^{2})$ and
$\displaystyle c^H_0 = \frac{5\varepsilon}{6} + O(\varepsilon^{2})$
while for $\varepsilon>0$,
$\displaystyle
c^K_0 = -\frac{2\varepsilon}{3} + O(\varepsilon^{2})$ and
$\displaystyle c^H_0 = -\varepsilon + O(\varepsilon^{2})$.

These results give the wrong continuum limit ($\varepsilon \rightarrow 0$)
for the FP determinant, both for positive and negative $\varepsilon$.

The problem is that the choice of the heat kernel cannot be made a
priori, but should be an outcome of the meaning of the tip of the cone
as a locus of infinite positive $(\alpha<1)$ or negative curvature
$(\alpha>1)$. We shall examine the problem by looking at the cone as a
limit case of a regular geometry. For positive curvature we shall
describe the tip of the cone as a segment of a sphere which connects
smoothly with the cone and for negative curvature we shall describe
the tip of the cone as a segment of the Poincar\'e pseudo-sphere of
constant negative curvature.

The limit we are interested in, is the one of the radius of the sphere
going to zero, keeping constant the integrated curvature.  The
sphere of radius ${1\over 2}\rho$, of constant curvature $R=-2
e^{-2\sigma} \Delta \sigma = 8\rho^{-2}$ or the pseudo-sphere of constant
curvature $R=-8\rho^{-2}$ are described by the conformal projection on
the complex $\omega$ plane where the metric is given by the conformal
factor $e^{2\sigma}=(1\pm u \bar u)^{-2}$ with $u=\omega/\rho$.

The integrated curvature on the segment of the sphere or pseudo-sphere
between $\tau=0$ and $\tau_0 = \rho v_{0}$, ($\tau = |\omega|$ and
$v=|u|$), is given by $\pm 8\pi v_{0}^{2}/(1\pm v_{0}^{2})$ and is
related to the opening angle of the cone $2\pi \alpha$ by
\begin{equation}
v_{0}^{2} ={1-\alpha \over 1+\alpha}~~{\rm ~for~ the~ sphere~~}
(0<\alpha<1)  \label{v0sfera}
\end{equation}
and
\begin{equation}
v_{0}^{2} ={\alpha-1  \over \alpha+1}~~{\rm ~for~ the~ pseudo-sphere~~}
(1<\alpha).   \label{v0pseudo}
\end{equation}

 From (\ref{v0sfera}) and (\ref{v0pseudo}) we see that a segment of
sphere or pseudo--sphere with given integrated curvature or
equivalently, given angular opening of the cone that matches with the
segment of the sphere, is described by a fixed value $v_{0}$,
corresponding to $\tau_{0} = \rho v_{0}$.

On the other hand the cone is described on the $\omega$ plane by the
transformation $z = x + i y = c \omega^{\alpha}$, which
produces on
the $\omega$ plane the conformal factor $e^{2\sigma} = c^2 (\omega
\bar{\omega})^{\alpha - 1}$ where $c$ is fixed by the relation
\begin{equation}
(1\pm \left(\frac{\tau_{0}}{\rho}\right)^{2})^{-2} = c^2
(\tau_{0}^{2})^{\alpha - 1},
\end{equation}
being $\tau_{0}$ the radius in the $\omega$--plane at which the sphere
or the pseudo-sphere connects to the cone. Hence we obtain $c =
\rho^{1-\alpha} \displaystyle{\frac{v_{0}^{1-\alpha}}{1 \pm
v_{0}^{2}}}$. In going over from the $z$ to the $\omega$--plane the vector
field $\xi$ acquire a Jacobian factor $c\bar{\omega}^{\alpha - 1}$ and the
tensor field $h$ a factor $(c\bar{\omega}^{\alpha - 1})^{2}$. We are
interested in the eigensolutions of $L^{\dag}L$ on such a regularized
cone which are regular for $\omega=0$. We shall compute these
eigenfunctions on the sphere or pseudo--sphere and then connect them
smoothly with those on the cone.

Solving explicitly the eigenvalue equation we find, for the eigensolutions
with orbital angular momentum $m$ on the sphere
\begin{eqnarray}
m=n\geq0 ~~~~~ & \xi^{(n)} =
\displaystyle{\frac{u^{n}}{(1+v^{2})^{2}}}  ~_{2}F_{1}
(\gamma_{1} + 2, 1 - \gamma_{1}; n + 1 ; \frac{v^{2}}{1+ v^{2}}) \\
m=-n\leq0 &  \xi^{(n)} = \bar{u}^{n} ~_{2}F_{1} (\gamma_{1}, -1
- \gamma_{1}; n+1; \displaystyle{\frac{v^{2}}{1+ v^{2}}})
\end{eqnarray}
where $\gamma_{1} = \frac{1}{2}(-1+\sqrt{9+4(\rho\mu)^{2}}~)$.
For $\rho^2 = 0$ they reduce to
\begin{eqnarray}
m = n \geq 0 &   \displaystyle{\frac{u^{n}}{ (1 +
    u \bar{u})^{2}}} \\
m= - n \leq 0 ~~~&  \displaystyle{\frac{\bar{u}^{n}}{ (1 +
u\bar{u})^{2}}}
[ (1 + u \bar{u} )^{2} - {2\over n+1} u \bar{u} ( 1 +
u \bar{u})^{2} + {2\over (n+1)(n+2)} (u \bar{u})^{2} ].
\end{eqnarray}
On the pseudo-sphere we have
\begin{eqnarray}
m=n\geq0 ~~~~~& \xi^{(n)} = \displaystyle{ \frac{u^{n}}{(1 -
    v^{2})^{2}}} ~_{2}F_{1}
(\gamma_{2} + 2, 1 - \gamma_{2}; n + 1 ; \frac{v^{2}}{v^{2} - 1} ) \\
m=-n\leq0 & \xi^{(n)} = \bar{u}^{n} ~_{2}F_{1} (\gamma_{2}, -1
-\gamma_{2}; n + 1; \displaystyle{\frac{v^{2}}{v^{2} - 1 }})
\end{eqnarray}
where $\gamma_{2} = \frac{1}{2}(-1+\sqrt{9-4(\rho\mu)^{2}}~)$.
For $\rho^2 = 0$ we obtain
\begin{eqnarray}
m = n \geq 0 &   \displaystyle{\frac{u^{n}}{ (1 - u \bar{u})^{2}}} \\
m=- n \leq 0 ~~~ &  \displaystyle{\frac{\bar{u}^{n}}{ (1 -
    u\bar{u})^{2}}}  [(1 - u\bar{u} )^{2} + {2\over n+1}
u\bar{u} ( 1 - u\bar{u})^{2} + {2 \over (n+1)(n+2)}
(u\bar{u})^{2} ].
\end{eqnarray}

The general eingensolution on the cone for orbital angular momentum $m$ has
the form
\begin{equation}
\xi_{\rm ext}^{(m)} = \left( \frac{u}{\bar{u}} \right)^{\frac{m}{2}} (u
  \bar{u})^{\frac{\alpha -1}{2}} \left[ a(\rho) J_{\gamma}
(2\rho\mu p v^{\alpha})  + b(\rho) J_{-\gamma} (2\rho\mu p v^{\alpha}) \right]
\end{equation}
where $\displaystyle{\gamma=\frac{m+\alpha -1}{\alpha}}$ and
$\displaystyle{p = \frac{v_{0}^{1-\alpha}}{1 \pm v_{0}^{2}}}$.
The coefficient $a(\rho)$ and $b(\rho)$ are fixed by
requiring the continuity of the logarithmic derivative of
$e^{-2\sigma} \xi$ with respect to
$\bar{\omega}$ at $\mid \! \omega\!\! \mid = \tau_{0}$.
In fact from the structure of eigenvalue equation $e^{-2\sigma}
\frac{\partial}{\partial\omega} e^{2\sigma}
\frac{\partial}{\partial\bar{\omega}} e^{- 2\sigma} \xi = - \mu^{2}
\xi$, we see that failing to satisfy such a condition would produce a
singular contribution at the matching point.
We are interested in the matching condition in the limit where the
regulator $\rho$ tends to 0.

Let us consider first $m=n\geq0$. For small $\rho$ the interior
solution multiplied by the factor $e^{-2\sigma}$ becomes
\begin{equation}
e^{-2\sigma} \xi_{int} = u^{n} \left[ 1 + (\rho\mu)^{2} f \left(
\frac{u\bar{u}}{1 \pm u\bar{u}}  \right) +
O((\rho\mu)^{4}) \right]
\end{equation}
while the exterior solution multiplied by the conformal factor
$e^{-2\sigma}$ becomes
\begin{equation}
  (\rho\mu)^{\gamma} a(\rho) u^{n} \left[ 1 + c_{1} (\rho\mu)^{2}
  (u\bar{u})^{\alpha} +O((\rho\mu)^{4}) \right] +
  (\rho\mu)^{-\gamma} b(\rho) \bar{u}^{-n} \left[  (u
  \bar{u})^{1-\alpha}  + O((\rho\mu)^2) \right]  \label{int1}.
\end{equation}

We notice that the lowest order in the first term of (\ref{int1}) has
vanishing derivative with respect to $\bar{\omega}$. Thus the
continuity of the logarithmic derivative for small $\rho\mu$ takes the form
\begin{equation}
  \frac{1}{\rho} k(\rho\mu)^{2} = \left. \frac{1}{\rho} \frac{a(\rho)
    c_{1} (\rho\mu)^{2} +
    b(\rho) c_{2} (\rho\mu)^{-2\gamma }}{ a(\rho) c_{3} + c_{4}
      b(\rho) (\rho\mu)^{-2\gamma} } \right.
\end{equation}
which solved in $\displaystyle{\frac{b(\rho)}{a(\rho)}}$ gives
\begin{equation}
  \left. \frac{b(\rho)}{a(\rho)} =  \frac{ (\rho\mu)^{2 + 2 \gamma} (c_{1}
    - k c_{3}) }{k (\rho\mu)^{2} c_{4} - c_{2}}  \right. .
\label{b1}
\end{equation}
Thus for $m\geq 0$ we see that for $2 + 2 \gamma > 0$, i.e.\ $\alpha >
\frac{1}{2}$, $b(\rho)$ vanishes when the regulator is removed at
constant integrated curvature.

Similarly one can deal with $m=-n<0$. In this case the derivative of
the interior solution multiplied by the conformal factor tends  to a
finite limits for $(\rho\mu)^{2} \rightarrow 0$ and the analog of
equation (\ref{b1}) is
\begin{equation}
\left. \frac{a(\rho)}{b(\rho)}  =  \frac{  (c_{5} - k_{1} c_{7})
  (\rho\mu)^{-2\gamma} }{k_{1} c_{8} - c_{6}}
  \label{b2} \right. .
\end{equation}
Thus for $m<0$ we have $a(\rho) \rightarrow 0$ for $\gamma<0$, i.e.\
for $\alpha<2.$

Thus we reached the conclusion that for the opening of the cone
$2\pi\alpha$ with $\frac{1}{2} < \alpha < 2$, as the regulator is
removed, only the term $J_{\frac{m+\alpha -1}{\alpha}}$ survives for
$m\geq 0$, while for $m<0$ the surviving term is $J_{-\frac{m+\alpha
-1}{\alpha}}$. Going over to the coordinate $z$, the heat kernel
$K(x,x';t)$ is thus given by (\ref{green1}) with $\delta = \alpha -1$.

We come now to the heat kernel $H$ for the field $h$. The
requirement \cite{alva} that $\det'(L^{\dag}L) = \det'(LL^{\dag})$
fixes the eigenfunctions of $LL^{\dag}$ to $h=L\xi$. Thus the
eigenfunctions of $LL^{\dag}$ are given in the $z$ coordinate for
$m\geq0$ by
\begin{equation}
  \left. \frac{\partial}{\partial\bar{z}} \left[ \left( \frac{z}{\bar{z}}
\right)^{\frac{\gamma}{2}} J_{\gamma} (2 \mu (z \bar{z})^{\frac{1}{2}})
\right] \right.
\end{equation}
which through a well known identity on the Bessel functions equals
\begin{equation}
  - \mu \left( \frac{z}{\bar{z}} \right)^{\frac{\gamma + 1}{2}}
  J_{\gamma + 1} (2 \mu (z \bar{z})^{\frac{1}{2}}),
\end{equation}
while for $m<0$ they are given by
\begin{equation}
  \mu \left( \frac{z}{\bar{z}} \right)^{\frac{\gamma + 1}{2}}
  J_{-\gamma - 1} (2 \mu (z \bar{z})^{\frac{1}{2}}),
 \end{equation}
always with $\displaystyle{\gamma = \frac{m + \alpha - 1}{\alpha}}$.
The net result is that the heat kernel $H$ is given by (\ref{green1})
with $\delta = 2\alpha -1$.

We come now back to eq.(\ref{var}). Applying Dowker's procedure to $ K =
K_{\alpha,\alpha-1}$ we obtain
\begin{equation}
  c_{0}^{K} =
  \frac{(\alpha - 1)(\alpha - 2)}{2\alpha} + \frac{1 - \alpha^{2}}{12 \alpha}
\label{zeta1}
\end{equation}
and for $H = K_{\alpha, 2 \alpha -1}$
\begin{equation}
   c_{0}^{H}=
  \frac{(2\alpha - 1)(2\alpha-2)}{2\alpha} + \frac{1 - \alpha^{2}}{12 \alpha}
\label{zeta2}.
\end{equation}
We notice that
$2 (c_{0}^{K} - c_{0}^{H}) = 3 (1-\alpha)$. Due to the local character
of the coefficients
of the asymptotic expansion of the trace of the heat kernel
\cite{dow}, for a generic compact surface without boundaries
such a relation becomes $3\sum_i(1-\alpha_i)$ where the sum runs over
the vertices and can be rewritten as $3\chi$ being $\chi$ the Euler
characteristic of the surface, in agreement with the Atiyah--Singer
\cite{atiy} index theorem applied to $L^{\dag}L$ and $LL^{\dag}$
\cite{alva}.  We
remark that such a topological relation is satisfied only by our
choice $\delta =
\alpha -1$ for $K$, among all the possibilities $\delta = \alpha -N$.

The terms concerning the zero modes in eq.(\ref{var}) in the continuum
approach are cancelled by the variations of other non--local terms
appearing in the measure \cite{moor}; so, we shall concentrate on the
$H$ and $K$ terms and  show that such terms can be explicitly
computed, giving a results which is 26 times the conformal anomaly of a
scalar field as it happens in the continuum.

Substituting the explicit value of $\delta \sigma$ eq.(\ref{delta}) in
equation (\ref{var}) and noting that $c_{0}^{K}$ and
$c_{0}^{H}$ depend only on $\alpha$ we have
\begin{eqnarray}
\delta{Z_{\alpha,\lambda}^{K}}' (0) = \gamma_{E} \delta c_{0}^{K}
+ (\delta \lambda -\lambda \frac{\delta \alpha}{\alpha}) [ 4
c_{0}^{K} - 2 c_{0}^{H}] + & \nonumber \\
+4 \frac{\delta \alpha}{\alpha} \int\!dx\, \ln (\alpha|x|) \, K(x,x,t)
-2 \frac{\delta \alpha}{\alpha} \int\!dx\, \ln (\alpha|x|) \, H(x,x,t).
\end{eqnarray}
Using (\ref{zeta1}) and (\ref{zeta2}) we obtain
\begin{eqnarray}
\delta{Z_{\alpha,\lambda}^{K}}' (0) = \gamma_{E} \delta c_{0}^{K}
+(\delta \lambda -\lambda \frac{\delta \alpha}{\alpha})
\displaystyle{\frac{13}{6}} (\frac{1}{\alpha}-\alpha) +
\frac{\delta\alpha}{\alpha}  \left[
4I_{K}(\alpha) -2 I_{H}(\alpha) \right] = & \nonumber \\
= \delta \left( \frac{13}{6} \lambda(\frac{1}{\alpha} - \alpha)
\right) + \delta \left[ \gamma_{E} \frac{(5\alpha -13)(\alpha - 1)}{12\alpha}
\right] +\frac{\delta\alpha}{\alpha} f(\alpha)
+ \frac{13}{3} \lambda \delta{\alpha}
\label{deltaz}
\end{eqnarray}
where the two integrals  $I_{K}(\alpha)$ and  $I_{H}(\alpha)$ have a
form very similar that of Aurell and Salomonson \cite{aur}.

Eq.(\ref{deltaz}) holds for a cone; the extension to a generic Regge
surface is done exactly along the lines of \cite{aur}.
If the surface is compact and closed we have $\sum_{i} \delta
\alpha_{i} = 0$ and the last term in (\ref{deltaz}) can be dropped.
Up to now we have considered $L^{\dag}L$ acting on $\xi$. Actually, as
the complex field has two independent components, one should consider
at the same time the complex conjugate operator $\overline{L^{\dag}L}$
acting on $\bar{\xi}$ \cite{alva}, with the result that all the traces
up to now computed have to be multiplied by 2.

Finally denoting with $w_{i}$ the complex coordinates of the vertices
on the plane on which the Regge surface is unfolded as done in
\cite{aur}, we obtain for the logarithm of the FP determinant
\begin{equation}
\frac{26}{6} \sum_{i,j\neq i}
\frac{(1-\alpha_{i})(1-\alpha_{j})}{\alpha_{i}}  \ln
|w_{i} - w_{j}| - \sum_{i} F(\alpha_{i}) \label{risultato}
\end{equation}
where $F(\alpha)$ is the primitive of $\displaystyle \frac{f(\alpha)}{\alpha}$
added to $\gamma_{E} c_{0}^{K}(\alpha)$ and can be written in the form of
an integral representation.

The term with the double sum is exactly 26 times the analogous term
appearing in the result of Aurell and Salomonson for the determinant
of a scalar field, i.e.\ the conformal anomaly. In the continuum limit
$\alpha_{i} - 1 \sim \frac{1}{N}$ being $N$ the number of vertices and
for $N \rightarrow \infty $ it goes over to the correct continuum limit
\cite{aur}.  $\sum_{i} F(\alpha_{i})$ in the continuum limit
($\alpha_{i} = 1 + \varepsilon_{i}, \varepsilon_{i} \rightarrow 0$)
goes over to $F'(1) \sum_{i} \varepsilon_{i} + \rm{cost.}$ and thus
becomes a topological term.  Generalizations to surfaces of higher
genus are easily obtained.

We point out that our results, derived by the regularization of the
tip of the cone by means of a segment of sphere or pseudo--sphere, are
largely independent of the details of this regularization. In fact the
effect of our regularization is that to impose (apart from a
correction that behaves like $\rho^{2}$ and that vanishes in the limit
$\rho \rightarrow 0$) a fixed logarithmic derivative of $e^{-2\sigma}
\xi$ at the boundary, combined with the fact the for $m\geq 0$ the
regular eigenfunction of $L^{\dag}L$ to the null eigenvalue has the
form $e^{2\sigma} \omega^{m}$.

Imposing Dirichlet boundary condition for the field $\xi$
on a small circle and then making the radius of the circle vanish,
reproduce eq.(\ref{risultato}) only for $1 < \alpha <2$.
At the same time the field $h=\frac{\partial}{\partial \bar{z}} \xi$
will violate the Dirichlet boundary condition in the same range of
$\alpha$, as $\delta = 2 \alpha - 1$ no longer lies in the limits $0<
\delta \leq 1$.
With Neumann boundary conditions we have the same situation in the
interval $\frac{1}{2} < \alpha <2$.

Our boundary conditions span the whole range $\frac{1}{2} < \alpha <2$, at
the boundary of which both the $L^{2}$ character of the eigenfunctions
and the procedure for obtaining eq.(\ref{dow}), for $\delta =
\alpha - 1$ or $\delta = 2\alpha -1$, are lost.
For $0<\alpha \leq \frac{1}{2}$ with our boundary conditions we obtain
$\delta = \alpha$ and for $N < \alpha \leq N+1$ we have $\delta =
\alpha -N$, thus introducing a non analytic behavior of ${\det}' L^{\dag}L$
as a function of $\alpha$.
This is not completely unexpected as a non analytic behavior is
already present in $c_{0}$ for the Green function of a particle on a
cone as a function of the magnetic flux through the tip of the cone
\cite{dow}, which in our case represents the phase change.

\end{document}